\documentstyle[twocolumn,aps,epsfig]{revtex}

\begin{document}

\title{Signatures in the Planck Regime}

\author{S.~Hossenfelder${}^a$\thanks{email: 
hossi@th.physik.uni-frankfurt.de}, 
M.~Bleicher${}^a$, S.~Hofmann${}^b$, J.~Ruppert${}^a$, 
S.~Scherer${}^a$, H.~St\"ocker${}^a$}

\address{${}^a$ Institut f\"ur Theoretische Physik\\ 
J. W. Goethe-Universit\"at\\
Robert-Mayer-Str. 8-10\\ 
60054 Frankfurt am Main, Germany}

\address{ ${}^b$ Department of Physics,\\ 
Stockholm University, \\
SCFAB, SE-106 91, \\
Stockholm, Sweden}

\maketitle

\noindent
\begin{abstract}
String theory suggests the existence of a minimum length scale. An exciting 
quantum mechanical implication of this feature is a modification of the
uncertainty principle.
In contrast to the conventional approach, this generalised uncertainty 
principle does not allow to resolve space time distances below the
Planck length. 
In models with extra dimensions, which are also motivated by string theory, 
the Planck scale can be lowered to values accessible by ultra high energetic cosmic
rays ({\sc UHECR}s) and by future 
colliders, i.e. $M_f\approx$ 1~TeV. 
It is demonstrated that in this novel scenario, short distance 
physics below $1/M_f$ is completely cloaked by the uncertainty principle. 
Therefore, Planckian effects could be the final physics discovery 
at future colliders and in {\sc UHECR}s.
As an application, we predict the modifications to the $e^+e^- \to f^+f^-$ cross-sections. 

\vspace{1cm}
\end{abstract}

\section{Introduction}

Even if a full description of quantum gravity is not yet available, there
are some general features that seem to go hand in hand with all promising candidates
for such a theory. One of them is the need for a higher dimensional space-time,
one other the existence of a minimal length scale. 
The scale at which the running couplings unify and quantum gravity
is likely to occur is called the Planck scale.
At this scale the quantum effects of gravitation get as important as those
of the electroweak and strong interactions.

In this paper we will implement both of these extensions in the standard model
without the aim to derive them from a fully consistent theory. Instead, we will 
to analyse some of the main features that may arise by the assumptions of 
extra dimensions and a minimal length scale.

In perturbative string theory \cite{Gross,Amati:1988tn},
the feature of a fundamental minimal length scale arises from the fact that strings can not probe 
distances smaller than the string scale. If the energy of a string reaches the Planck mass 
$m_{\rm p}$, excitations of the string can occur and cause a non-zero extension \cite{witten}. 
Due to this, uncertainty in position measurement can never become smaller than 
$l_{\rm p}=\hbar/m_{\rm p}$. For a review, see \cite{Garay:1994en,Kempf:1998gk}.
   
Naturally, this minimum length uncertainty is related to a modification 
of the standard commutation relations between position and momentum \cite{Kempf:1994su,Kempf:1996nk}. Application of this is of high
interest for quantum fluctuations in the early universe and for inflation \cite{Hassan,Danielsson:2002kx,Shankaranarayanan:2002ax,Mersini:2001su,Kempf:2000ac,Kempf:2001fa,Martin:2000xs,Easther:2001fz,Brandenberger:2000wr}. 

The incorporation of the modified commutation relations into quantum theory is not
fully consistent in all approaches, therefore we will define physical variables 
step by step. 

The existence of a minimal length scale becomes important even for collider physics
with the further incorporation of the central idea of Large eXtra Dimensions ({\sc LXD}s).
The model of {\sc LXD}s, which was recently
proposed in \cite{add1,add2,add3,Randall:1999vf,Randall:1999ee}, might 
allow to study interactions
at Planckian energies in the next generation collider experiments. 
Here, the hierarchy-problem is solved
or at least reformulated in a geometric language by the existence of $d$ 
compactified {\sc LXD}s
in which only the gravitons can propagate. The standard model particles 
are bound to our 4-dimensional
sub-manifold, often called our 3-brane. This results in a lowering of the Planck scale to a new
fundamental scale, $M_f$, and gives rise to the exciting possibility of TeV scale {\sc GUT}s \cite{Dienes}.

The strength of a force at a distance $r$ generated by a charge depends 
on the
number of space-like dimensions. For distances smaller than the 
compactification radius $R$,
the gravitational interaction drops faster compared to the other 
interactions.
For distances much bigger than $R$, gravity is described by the 
well known potential
law $\propto 1/r$. However, for $r \le R$ the force lines 
are diluted into the extra dimensions. Assuming a smooth transition to 
Newton's law, this results in a smaller effective coupling 
constant for gravity.

This leads to the following relation between the four-dimensional 
Planck mass $m_{\rm p}$ and
the higher dimensional Planck mass $M_f$, which is the new 
fundamental scale of the theory:
\begin{eqnarray} \label{master}
m_{\rm p}^2 = R^d M_f^{d+2} \quad.
\end{eqnarray}

The lowered fundamental scale would lead to a vast number 
of observable phenomena 
for quantum gravity at
energies in the range $M_f$. In fact, the non-observation of these predicted
 features gives
first constraints on the parameters of the model, the number of extra
 dimensions $d$ and the fundamental scale
$M_f$ \cite{Revcon,Cheung,Cullen:2000ef}. On the one hand, this scenario has major 
consequences for cosmology and
astrophysics  such as the modification of inflation in the early 
universe and enhanced
supernova-cooling due to graviton emission \cite{add3,Cullen,Probes,astrocon,GOD}.
On the other hand, additional processes are expected in high-energy 
collisions \cite{Rums}: production of real and
virtual gravitons \cite{enloss1,enloss2,Hewett,Nussi,Rizzo} and the
 creation of black holes at
energies that can be achieved at colliders in the near future \cite{adm,Giddings:2001ih,Mocioiu:2003gi,Kotwal:2002wg,Uehara:2001yk,Emparan:2001kf,Hossenfelder:2001dn} and in ultra high energetic cosmic rays \cite{Ringwald:2001vk,Kazanas:2001ep}.

This paper is organised as follows. We will begin with a 
sketch of the basics of quantum mechanics (section II), and in the third section  
modify these familiar relations by introducing generalised uncertainty.
This will be done in $1+1$ dimensions first, then we care for
the full $3+1$ dimensional description (this is understood to be the analysis on our brane).
To examine the phenomenological implications on a basic level,
we first analyse the modified
Schr\"odinger Equation, 
the Dirac Equation and the Klein-Gordon Equation in sections IV-VI. 
In section VII we investigate the influence 
of the minimal length scale on {\sc QED} cross-sections at tree-level and compare with $e^+e^-$ data. 
Section VIII provides an
estimation of the effect on graviton production. We end with a conclusion of our results
in section IX.

In the following, we use the convention $\hbar = L_f M_f$, $c=1$. 
Greek indices $\alpha,\mu,...$ run from 0 to 3. Latin indices $i,j,...$ run from 1 to 3, 
latin indices $a,b,...$ run from 4 to $4+d$. 
In order to distinguish the ordinary quantities (e.g.~$E$) from the modified ones,
we label the latter with a tilde ($\tilde{E}$).

\section{The Uncertainty Relation}

In standard quantum mechanics 
 translations in space and time are generated by 
momentum $p_i$ and energy $E$, respectively. However, from
purely dimensional reasons, the generators of the translations in space and time are the
wave vector $k_i$ and the frequency $\omega$. The relation between $(k_i,\omega)$ and 
$(p_i,E)$ is usually given, of course, by the constant $\hbar$ (often chosen to be equal one):

\begin{eqnarray}
p_i&=& \hbar k_i\quad,\\
E &=& \hbar \omega\quad.
\end{eqnarray}
In the present context it is of utmost importance to re-investigate this relation
carefully.

Using the well known commutation relations 
\begin{eqnarray} \label{CommXK}
[\hat x_i,\hat k_j]={\mathrm i } \delta_{ij}\quad,
\end{eqnarray}
quantisation in position representation $\hat{x}_i = x_i$ leads to: 
\begin{eqnarray} 
\hat{k}_i = - {\mathrm i} \partial_i \quad&,&\quad
\hat{p}_i = \hbar \hat{k}_i  = - {\mathrm i} \hbar \partial_i \quad, \\
\hat{\omega} =  + {\mathrm i} \partial_t \quad&,&\quad
\hat{E}      =  \hbar \hat{\omega} \; =  + {\mathrm i} \hbar \partial_t\quad.
\end{eqnarray}

In the momentum representation, $\hat{p}_i = 
p_i$, 
the commutation relation is fulfilled by 
\begin{eqnarray} \label{MomentumRepresentation}
\hat{x}_i = {\mathrm i} \hbar \frac{\partial}{\partial p_i} = {\mathrm i} \frac{\partial p_i}{\partial k_i} \frac{\partial}{\partial p_i} = {\mathrm i} \frac{\partial}{\partial k_i}\quad.
\end{eqnarray}
The general relation for the root mean square deviations for the expectation values of
two operators $\hat{A}$ and $\hat{B}$,
\begin{eqnarray} 
\Delta A \Delta B \geq \frac{1}{2} \left|\langle[\hat{A},\hat{B}]\rangle\right|\quad,
\end{eqnarray}
then leads to the uncertainty relation
\begin{eqnarray} 
\Delta p_i \Delta x_i \geq \frac{1}{2} \hbar\quad.
\end{eqnarray}
The equation of motion (no explicit time dependence) for the wave function is generated by the evolution operator $\hat{U}$:
\begin{eqnarray} 
\vert \psi(t)\rangle &=& \hat{U}(t-t_o) \vert \psi(t_o)\rangle\quad,\\
\hat{U}(t-t_o) &=& \exp \left( -\frac{{\rm i}}{\hbar} \hat{E} (t-t_o) \right)\quad,\\
\Rightarrow\quad + {\rm i} \hbar \partial_t \vert \psi \rangle &=& \hat{E} \vert \psi \rangle\quad.
\end{eqnarray}
The time dependence of an operator $\hat{A}$ (no explicit time dependence) is (in the Heisenberg picture) then given by   
\begin{eqnarray} \label{timeevo}
\frac{{\rm d} }{{\rm d}t} \hat{A} = [\hat{A},\hat{E}] \quad.
\end{eqnarray}

\section{Generalised Uncertainty}

In order to implement the notion of a minimal length $L_f$, 
let us now suppose that one can increase $p$ arbitrarily, but that $k$ has an upper bound. 
This effect will show up when $p$ approaches a certain scale $M_f$. 
The physical interpretation of this is that particles could not possess arbitrarily
small Compton wavelengths $\lambda = 2\pi/k$ and that arbitrarily small scales
could not be resolved anymore. 

To incorporate this behaviour, we assume a relation $k=k(p)$ between $p$ and $k$ 
which is an uneven function (because of parity) and which asymptotically
approaches $1/L_f$.\footnote{Note that this is similar to introducing an energy dependence of Planck's
constant $\hbar \to \hbar(p)$.} Furthermore, we demand the functional relation between the  energy $E$ 
and the frequency $\omega$ to be the same as that between the wave vector $k$ and the momentum $p$.

In contrast to \cite{Hassan}, there is no modified dispersion relation in our approach, 
since $\partial \omega/\partial k = \partial E / \partial p$. This means 
that the functional behaviour of $k(p)$ is the same as that of $\omega(E)$ 
up to a constant. A possible choice for these relations is

\begin{eqnarray}
L_f k(p) &=& \tanh^{1/\gamma} \left[ \left( \frac{p}{M_f} 
\right)^{\gamma} \right] \quad,\\
L_f \omega(E) &=& \tanh^{1/\gamma} \left[ \left( \frac{E}{M_f} \right)^{\gamma} 
\right]\quad,
\end{eqnarray}
with a real, positive constant $\gamma$. For simplicity, we will use $\gamma = 1$.

In the following we will study two approximations, from here on referred to as cases (a) and (b):
\begin{itemize}
\item[(a)]
The regime of first effects including order $(p/M_f)^3$ contributions and 
\item[(b)] 
The high energy limit $p \gg M_f$. 
\end{itemize}
Expanding $\tanh(x)$ for small arguments gives for case (a)
\begin{eqnarray}
L_f k(p) &\approx& \frac{p}{M_f} - \frac{ 1}{3} 
\left( \frac{p}{M_f} \right)^3 \label{app1a}\\
L_f \omega(E) &\approx& \frac{E}{M_f} - \frac{ 1}{3} \left( \frac{E}{M_f} \right)^3\\
\frac{1}{M_f} p(k) &\approx& k L_f + \frac{ 1}{3} \left( k L_f \right)^3\\
\frac{1}{M_f} E(\omega) &\approx& \omega L_f 
+ \frac{ 1}{3} \left( \omega L_f \right)^3 \label{app4a} \quad.
\end{eqnarray}
This yields to 3$^{\rm rd}$ order
\begin{eqnarray}
\hbar \frac{\partial k}{\partial p} &\approx& 1 - \left( \frac{p}{M_f} \right)^2 \nonumber\\
&\approx& 1 - \left( k L_f \right)^2 \label{diff1a}\\
\frac{1}{\hbar} \frac{\partial p}{\partial k} &\approx& 1 + \left( k L_f \right)^2 \nonumber\\
&\approx& 1 + \left(\frac{p}{M_f} \right)^2 \label{diff2a}\quad.
\end{eqnarray}
In case (b) we have $\tanh(x)\approx \pm 1 \mp 2 \exp(\mp 2x)$ 
for \linebreak $\vert x \vert \gg1$, with the upper signs for positive values of $x$. 
Skipping one factor $2$ in the exponent, which can be absorbed by a redefinition of $M_f$, one obtains:
\begin{eqnarray}
L_f k(p) &\approx& \pm 1 \mp 2\exp\left(\mp \frac{p}{M_f}\right) \label{app1b}\\
L_f \omega(E) &\approx& \pm 1 \mp 2\exp\left(\mp \frac{E}{M_f}\right) \label{app2b} \\
\frac{1}{M_f} p(k) &\approx& \mp \ln \left(\frac{1 \mp k L_f}{2} \right) \\
\frac{1}{M_f} E(\omega) &\approx& \mp \ln \left(\frac{1 \mp \omega L_f}{2} \right) \label{app4b} \quad.
\end{eqnarray}
The derivatives are
\begin{eqnarray}
\hbar \frac{\partial k}{\partial p} &\approx& 2\exp \left( - \frac{\vert p \vert}{M_f}\right) \label{diff1b}\\ 
\frac{1}{\hbar} \frac{\partial p}{\partial k} &\approx& \frac{1}{2}\frac{1}{1 \mp k L_f } 
\;  .
\label{diff2b} 
\end{eqnarray}
The quantisation of these relations is straight forward. The commutators between $\hat{k}$ and $\hat{x}$ 
remain in the standard form given by Eq.~(\ref{CommXK}). Inserting the functional relation between the
wave vector and the momentum then yields the modified commutator for the momentum. 
With the commutator relation
\footnote{
Here, $\hat{A}$ is an operator valued polynom or formal series in $\hat{k}$.
The derivative on the right hand side has to be taken with respect 
to $k$ and then to be quantised.}  
\begin{eqnarray}
[\,\hat{x}, \hat{A}(k)] = + {\rm i} 
\frac{\partial A}{\partial k} \quad,
\end{eqnarray}
the modified commutator algebra  now reads
\begin{eqnarray}
[\,\hat{x},\hat{p}]&=& + {\rm i} \frac{\partial p}{\partial k} \quad.
\end{eqnarray} 
This results in the generalised uncertainty relation
\begin{eqnarray} 
\Delta p \Delta x \geq \frac{1}{2}  \Bigg| \left\langle \frac{\partial p}{\partial k} \right\rangle \Bigg| \quad. 
\end{eqnarray}

In case (a), with the approximations (\ref{app1a})-(\ref{app4a}), 
the results of Ref. \cite{Hassan} are reproduced:  
\begin{eqnarray}
[\hat{x},\hat{p}] 
&\approx& {\rm i} \hbar\left( 1 + \frac{ \hat{p}^2}{M_f^2} \right) 
\end{eqnarray}
giving the generalised uncertainty relation
\begin{eqnarray} 
\Delta p \Delta x \geq \frac{1}{2} \hbar \left( 1+ \frac{\langle \hat{p}^2 \rangle }{M_f^2} \right) \quad.
\end{eqnarray}

In the asymptotic case (b) this yields
\begin{eqnarray}
[\,\hat{x},\hat{p}] &\approx& {\rm i} \frac{\hbar}{2}\, \exp 
\left(+ \frac{|\hat{p}|}{M_f}\right)\quad, \\
\Delta p \Delta x &\geq& \frac{1}{4} \hbar\, \left\langle \exp \left(+ \frac{|\hat{p}|}{M_f}\right) \right\rangle \quad.
\end{eqnarray}

Quantisation proceeds in the usual   way from the commutation relations. 
For scattering theory it is convenient to work in the momentum representation, $\hat{p}=p, \hat{k}= k(p)$. 
From Eq.~\ref{MomentumRepresentation},
\begin{eqnarray} 
\hat{x}&=&{\rm i} \partial_k = {\rm i} \frac{\partial p}{\partial k} \partial_p 
\end{eqnarray}
we obtain in case (a) (first derived in Ref.~\cite{Kempf:1994su}):
\begin{eqnarray} \label{XinMomentum}
\hat{x} &\approx& {\rm i} \hbar \left( 1+ \frac{p^2}{M_f^2} \right)\partial_p\quad,
\end{eqnarray}
and in case (b):
\begin{eqnarray}
\hat{x} &\approx& {\rm i} \frac{\hbar}{2} \,\exp \left( \frac{\vert p \vert}{M_f}\right) \partial_p
\end{eqnarray}

As a first application of this approach to quantum mechanics, we will study the 
Schr\"odinger Equation  in section IV. 
Focusing on conservative potentials in non relativistic quantum mechanics
we give the operators in the position representation which is  
suited best for this purpose:
\begin{eqnarray}
\hat{x}&=&x\quad,\quad \hat{k}= - {\rm i} \partial_x\nonumber\\
\hat{p}&=& \hat{p}(\hat{k}) \quad,  
\end{eqnarray}
yielding in case (a)
\begin{eqnarray}
\hat{p} &\approx& - {\rm i} \hbar \left( 1 - \frac{L_f^2}{3}  \partial_x^2 \right)\partial_x \quad.
\end{eqnarray}
The new momentum operator now includes higher derivatives. 

Since $k=k(p)$ we have $\hat{p}(\hat{k})\vert k \rangle= p(k)\vert k \rangle$ and 
so $\vert k \rangle \propto \vert p(k) \rangle$. We could now add that both sets 
of eigenvectors have to be a complete orthonormal system and 
therefore $\langle k' \vert k \rangle = \delta(k-k')$, 
$\langle p' \vert p \rangle = \delta(p-p')$. 
This seems to be a reasonable choice at 
first sight, since  $\vert k \rangle$ is known from the cis-Planckian regime. 
Unfortunately, now the normalisation of the states is different 
because $k$ is restricted to the Brillouin zone\footnote{
We borrow this expression from solid state physics where
an analogous bound is present.}
$-1/L_f$ to $1/L_f$. 

To avoid the need to recalculate normalisation factors, we  
choose the $\vert p(k) \rangle$ to be identical to 
the $\vert k \rangle$. Following the proposal of \cite{Kempf:1994su} this yields
\begin{eqnarray}
\langle p' \vert p \rangle &=& \langle k(p) \vert k(p') \rangle = \delta\left(k(p)-k(p')\right)\nonumber \\
&=& \frac{\partial p}{\partial k} \delta\left(p-p'\right)
\end{eqnarray}
and avoids a new normalisation of 
the eigen functions by a redefinition of the measure in momentum space 
\begin{eqnarray}
{\mathrm d} p \rightarrow \frac{{\mathrm d} p}{\hbar}  \frac{\partial k}{\partial p} \quad.
\end{eqnarray}
This redefinition has a physical interpretation because we expect the momentum 
space to be squeezed at high momentum values and weighted less. 

For the different cases under discussion, one gets:\\
Case (a):
\begin{eqnarray}
\frac{{\mathrm d} p}{\hbar} \rightarrow \frac{{\mathrm d} p}{\hbar} \frac{1}{1+ (p/M_f)^2}\quad.
\end{eqnarray}
Case (b):
\begin{eqnarray}
\frac{{\mathrm d} p}{\hbar} \rightarrow \frac{{\mathrm d} p}{\hbar} \,2 \exp \left( - \frac{\vert p \vert}{M_f}\right)\quad.
\end{eqnarray}

The operator of time translation is no longer identical to the energy operator 
times $\hbar$ in this context. In ordinary quantum mechanics, both of them are 
$\propto \hat{H}$. To avoid confusion, let $\hat{\omega}$ be that operator defined
by the generator of the Lorentz-Algebra which belongs to the time translation 
and $\hat{E}=\hat{E}(\hat{\omega})$ the energy-operator for the free particle. 
$\hat{E}_{{\rm tot}} =\hat{E}(\hat{\omega})+V(\hat{x})$ is then the operator 
of the total energy, including a time-independent potential $V(x)$. 
The equation of motion for the wave function is then given by 
\begin{eqnarray} 
\hat{U}(t-t_o) &=& \exp \left( - {\rm i} 
\hat{\omega}(\hat{E}_{{\rm tot}}) (t-t_o) \right) \quad,\\
\Rightarrow  \quad {\rm i} \partial_t \vert \psi \rangle &=& 
 \hat{\omega}(\hat{E}_{{\rm tot}}) \vert \psi \rangle \label{EOM}\quad,
\end{eqnarray}
which has in case (a) the explicit form
\begin{eqnarray} 
{\rm i} \hbar \partial_t \vert \psi \rangle &\approx& 
+\left( \hat{E}_{{\rm tot}} - \hat{E}_{{\rm tot}}^3/3 M_f^2 \right) 
\vert \psi \rangle \quad.
\end{eqnarray}

\subsection{Lorentz Invariance and Conservation Laws in Four Dimensions} 
\label{lorentz}
We will use the following short notations: 

\begin{eqnarray} 
k&=&\vert \vec{k} \vert\quad ,\quad \vec{k}=(k_x,k_y,k_z)\quad,\quad{\underline{ \bf k}}=(\vec{k},\omega)\quad,\\ 
p&=&\vert \vec{p} \vert\quad ,\quad\vec{p}=(p_x,p_y,p_z)\quad,\quad {\underline{ \bf p}}=(\vec{p},E)\quad.
\end{eqnarray}

As discussed above, we leave the dispersion relation unmodified. 
However, as $E=\sqrt{p^2+m^2}$ expresses 
the relativistic energy-momentum relation we meet a serious problem at this point. 
The mass-shell relation is a consequence of ${\underline{ \bf p}}$ being a 
Lorentz vector rather than ${\underline{ \bf k}}$. 
Thus, we have to reconsider  Lorentz covariance in the 
trans Planckian regime.  For energy scales below $M_f$, an observer boosted to high velocities 
would observe  arbitrarily large energies. We have to assure then, that
the Lorentz-transformed ${\underline{ \bf k}}$ always stays below the new limit, which
means its transformation properties are not identical to those of the momentum
${\underline{ \bf p}}$. To put this in other words, a Lorentz boosted observer is not
allowed to see the minimal length further contracted.
Several proposals have been 
made to solve this problem. Most of them suggest a modification of the 
Lorentz transformation \cite{Amelino-Camelia:2002wr,Magueijo:2001cr,Toller:2003tz,Rovelli:2002vp}, but the treatment is still under debate. 

However, the appearance of this problem might not be as astonishing as it 
seems at first sight. 
Because the modifications we  examine do
occur at energies at which quantum gravity will get important, curvature 
corrections to the space time must  not be neglected anymore. 
Therefore, the quantities should be general relativistic 
covariant rather than flat space Lorentz covariant. 
These effects will then exhibit themselves in strong background fields, 
but here also the particle's curvature itself makes an essential contribution. The 
exact -- but unknown -- transformation should assure that no 
coordinate transformation can
push ${\underline{ \bf k}} $ beyond the Planck scale. For practical use of the 
modified quantum theory considered here, we treat the momentum
${\underline{ \bf p}}$ as the Lorentz covariant partner of the wave vector ${\underline{ \bf k}}$. We will
assume that  the momentum is Lorentz covariant and that the functional relation 
between the two quantities, although unknown, is of the desired behaviour. 
Inserting one of the approximations
at the end of the computation then breaks Lorentz covariance.
 
In fact, in the present scenario ${\underline{ \bf k}}$ is also not a conserved 
quantity in interactions, because the relation between ${\underline{ \bf p}}$ and
${\underline{ \bf k}} $ is not linear anymore. 
In single particle dynamics we have ,in generalisation of Eq. (\ref{timeevo}), 
the time evolution of the operator $\hat{A}$
\begin{eqnarray} 
\frac{{\rm d} }{{\rm d}t} \hat{A} = [\hat{A},\hat{\omega}]
\end{eqnarray}

Since $[\hat{A},\hat{B}] = 0$ is equivalent to $[\hat{A},f(\hat{B})]=0$ for 
any well defined functions $f$ of $\hat{B}$, quantities 
conserved in ordinary quantum mechanics are also conserved in the approach
considered here. In particular, the single particle momentum $\hat{p}$ 
and energy $\hat{E}$ are conserved if no interactions occur.

The canonical commutation relations are given by
\begin{eqnarray}  \label{commd4}
[\,x^{\nu}, p_{\mu}] &=& +{\rm i}  \frac{\partial p_{\mu}}{\partial k_{\nu}}\quad,
\end{eqnarray}
and
\begin{eqnarray}  
[p_{\nu},p_{\mu}] = 0 \quad,
\end{eqnarray}
with ${\underline {\bf p}}={\underline{ \bf p}}({\underline {\bf k}})$ 
being a Lorentz vector and fulfilling all requirements mentioned above. 

The invariant volume element is then modified to be 
\begin{eqnarray}
{\mathrm d}^4 p  &\rightarrow& \frac{{\mathrm d}^4 p}{\hbar^4}  \;{\rm det} \left( \frac{\partial k_{\mu}}{\partial p_{\nu}}\right) \nonumber \\
&=& \frac{{\mathrm d}^4 p}{\hbar^4} \prod_{\nu} \frac{\partial k_{\nu}}{\partial p_{\nu}}  \quad.
\end{eqnarray}
In the last step we used the rotational invariance of the relations $p_{\mu}=p_{\mu}(k_\mu)$. Due to this 
the Jacobi matrix is diagonal.

\subsection{The Non-Relativistic Case: Three Dimensions}

This generalises our $1+1$-dimensional treatment from the first section. Note that
this approach is no longer Lorentz invariant and therefore suited for non-relativistic
scenarios only.

\begin{eqnarray}  \label{commd3}
[\,\hat{x}_i, \hat{p}_j] & = & {\rm i} \hbar \delta_{ij} 
\frac{\partial p_{i}}{\partial k_{j}}\quad,\\{}
[\hat{p}_i,\hat{p}_j] & = & 0\quad.
\end{eqnarray}
Rotational invariance implies for case (a)

\begin{eqnarray}  \label{commd3a}
[\hat{x}_i, \hat{p}_j] & = & {\rm i} \hbar \delta_{ij} 
\left( 1+ \frac{\hat{p}^{\;2}}{M_f^2} \right)\quad,\\{}
[\hat{p}^i,\hat{p}_j] & = & 0\quad.
\end{eqnarray}
In the position representation $\vec{\nabla}=(\partial_x, \partial_y, \partial_z)$ the momentum operator then reads
\begin{eqnarray}  
\hat{\vec{p}} = -{\rm i} \hbar \vec{\nabla} \left(1 - \frac{L_f^2}{3} \vec{\nabla}^{\; 2} \right)
\end{eqnarray}

\section{Schr\"odinger Equation}

First we will have a look at the free scalar particle in the low energy limit. 
We will define physical variables step by step since different 
approaches to  incorporate  the minimal length  into quantum theory have been
given in the literature. 

Let us consider the modified Schr\"odinger equation. 
For usual one  gets it by quantising the low energy expansion, $p/m\ll1$, of the relativistic expression
\begin{eqnarray}
E=m\sqrt{1+\frac{p^2}{m^2}} = m \left(1 + \frac{p^2}{2m^2} + {\cal{O}}((p/m)^3) \right)
\end{eqnarray}
and dropping the constant term $m$ because an additive constant in the Hamiltonian 
does not change the dynamics. By multiplying a phase $\exp(-{\rm i} m t)$ to $\vert \psi \rangle$ 
we could get rid of it. But now this prescription is not applicable anymore because an additive constant in $E$ 
does not yield an additive constant in $\omega$ and therefore is has to be kept.  

With 
\begin{eqnarray}
E^3 = m^3 \left(1+ \frac{3p^2}{2m^2} + {\cal{O}}((p/m)^3) \right)
\end{eqnarray} 
the modified Schr\"odinger Equation, see (\ref{EOM}), is then given by
\begin{eqnarray} \label{freesch}
{\rm i} \hbar \; \partial_t \vert \psi \rangle &=& 
\left[ m \left( 1 - \frac{m^2}{3M_f^2} \right) + \frac{\hat{p}^2}{2m} \left( 1 - \frac{m^2}{M_f^2} \right) \right] \vert \psi \rangle \quad.
\end{eqnarray} 
The first term can be dropped again, since it contributes only an overall phase factor. 
This means, that up to order $p^2/M_f^2$ and $p^2/m^2$ no change in the dynamics occurs. 
However, the kept term $m$ will yield extra terms in higher order approximations. 
Equation (\ref{freesch}) will modify the frequency spectrum of very 
heavy ($m\approx M_f$) non relativistic particles and has therefore little applications. 

Fortunately,  we are mainly interested in general  in the energy spectrum 
and do not need to calculate $\hat{\omega}$ at all. Let us proceed now with the Schr\"odinger 
equation for a particle in a potential $V(\hat{x})$ with the two most prominent cases: 
the harmonic oscillator and the hydrogen atom. We want to calculate the modified energy 
levels $\tilde{E}_n$ as solutions of the time-independent Schr\"odinger equation. 
In the following we add $m\ll M_f$. The time dependence is split off by a separation 
of the variables and has the form $\exp(- {\rm i}\omega_n t)$ with $\omega_n=\omega(\tilde{E}_n)$.
  
\begin{eqnarray} \label{Eigenwert}
\tilde{E}_n \vert \psi \rangle &=& 
\left( \frac{\hat{p}^2}{2m} + V(\hat{x}) \right) \vert \psi \rangle \quad.
\end{eqnarray} 

For the harmonic oscillator with $V(x)= m\Omega^2 x^2 /2 $ in the momentum representation, 
we find using Eq.~(\ref{XinMomentum})
\begin{eqnarray} \label{HarmOsziMomentum} 
\tilde{E}_n \vert \psi \rangle &=& 
\left( \frac{p^2}{2m} - \frac{m\hbar^2\Omega^2}{2} \left( (1+p^2/M_f^2)\partial_p \right)^2 \right) \vert \psi \rangle \quad.
\end{eqnarray}
An analytic solution of this differential equation has been given in \cite{Kempf:1994su} and, 
for a more general setting, in \cite{Dadic}.

\begin{figure}[h]
\vskip 0mm
\vspace*{-0.5cm}
\centerline{
\psfig{figure=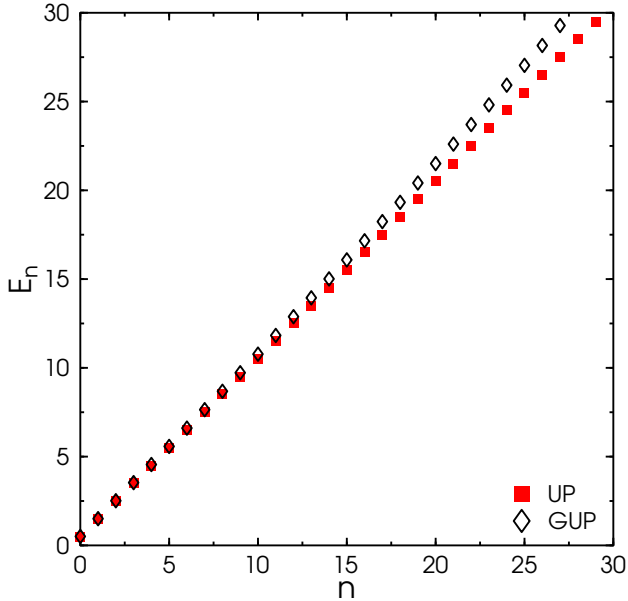,width=4.0in}}
\vskip 0mm
\caption{The energy levels of the harmonic oscillator with usual uncertainty (UP) in comparison to the case with 
generalised uncertainty (GUP). Here, $\Omega=m=1, M_f=0.1$.
\label{fig1}}
\end{figure}
The momentum space equation (\ref{HarmOsziMomentum}) is well suited 
for a numerical treatment. We have solved this eigenvalue problem numerically
 and it fits the analytically obtained values of \cite{Kempf:1994su} to very
high precision. 
This is plotted in Fig.~\ref{fig1} for $\Omega=m=1$ and $M_f=0.1$.
It can be seen qualitatively that the levels get shifted to higher energies with increasing $n$ in comparison to the usual $E_n \propto 1/2+n$.
If one tries to solve the eigenvalue equation in the position representation,
\begin{eqnarray} \label{HarmOsziPosition} 
\tilde{E}_n \vert \psi \rangle &=& -\frac{\hbar^2}{2m}\left[\left(1-\frac{L_f^2}{3}\partial^2_x\right)\,\partial_x\right]^2 + \frac{m\omega^2}{2}x^2  \vert \psi \rangle  \quad,
\end{eqnarray}
one has to cope with the higher derivatives, and adequate numerical methods to 
treat this stiff problem properly are quite involved.  
However, for practical purposes, one can consider the higher order terms 
as perturbations to the standard quantum mechanics problem and 
resort to perturbation theory, as was done analytically 
for the three dimensional harmonic oscillator in \cite{Kempf:1996fz}.
Our  perturbative calculations have shown numerically, that first order 
perturbation theory reaches 5 percent of the exact
eigenvalues of the harmonic oscillator in the {\sc GUP} regime. 

The hydrogen atom is treated best in position representation to avoid the 
problem of substituting $\hat{x} = {\rm i}\partial_k$ in the $1/r$ potential\footnote{It should be noted at
this point that in \cite{Akhoury:2003kc}, the hydrogen atom is treated with a minimal length uncertainty relation
in the momentum representation. However, in contrast to our approach, the authors of \cite{Akhoury:2003kc} use a 
modification of standard quantum mechanics where the new position operators do not commute anymore, 
$[\hat{x}_i,\hat{x}_j] \ne 0$. This prohibits the use of the position representation. Contrary to the concordant 
results presented in \cite{Brau}, and in this paper, the energy levels of the hydrogen atom are shifted downwards
in the approach of \cite{Akhoury:2003kc}.}.
To derive the equation for the Coulomb potential $V(r) = e^2 /r$ we will as usual first transform 
into spherical coordinates $r,\theta,\phi$ with $r=\vert \vec{r} \vert$. We look only at the case of 
vanishing angular dependence, $l=0$. (For a treatment of the full angular dependence see \cite{Brau}, 
who uses the perturbation theory method to calculate the shift in the energy spectrum). We
have then in position representation
\begin{eqnarray}  
\hat{p}_r= {\rm i} \hbar \frac{1}{r} \partial_r \left( 1 -  \frac{L_f^2}{3} \partial^2_r  \right) r\quad,
\end{eqnarray}
and for the energy operator we find
\begin{eqnarray}  
\hat{E} &=& 
- \frac{\hbar^2}{2m r}    \left(\partial_r -   \frac{L_f^2}{3} \partial^3_r   \right)^2 r + \frac{e^2}{r} \quad.
\end{eqnarray}
For the calculation of the eigenvalues $\tilde{E}_n$ of $\hat{E}$ we can substitute as usual 
$\vert \phi \rangle = r \vert \psi \rangle$ and then deal with the equation 
\begin{eqnarray}  
\tilde{E}_n \vert \phi \rangle &=& 
\left( - \frac{\hbar^2}{2m}    \left(\partial_r - \frac{L_f^2}{3} \partial^3_r   \right)^2  + \frac{e^2}{r}  \right) \vert \phi \rangle \quad.
\end{eqnarray}
As in the case of the harmonic oscillator, the higher derivatives can be treated as 
perturbations, and the
corresponding shifts of the energy levels can be calculated. 

There is, however, a second way to calculate the energy levels, which applies the semi classical 
calculation of Bohr to the generalised uncertainty principle.

\begin{figure}[h]
\vskip 0mm
\vspace*{-1cm}
\centerline{\psfig{figure=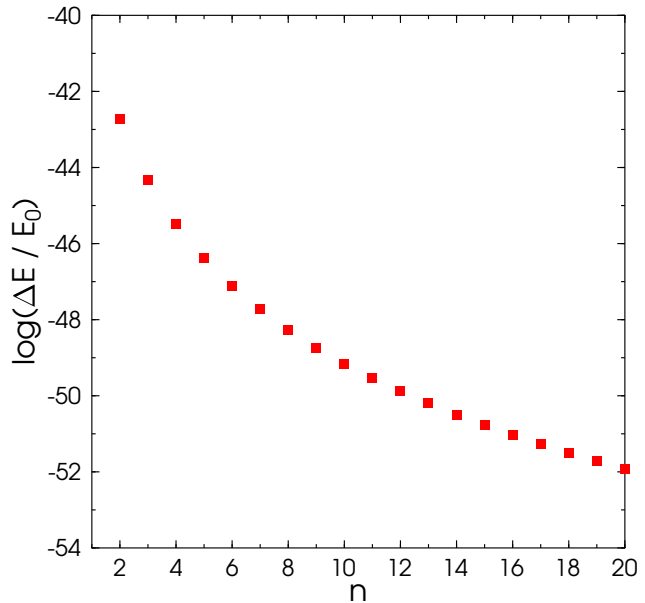,width=4.0in}}
\caption{The difference of energy levels $\Delta E =  \tilde{E}_n - E_n$ with $l=0$ in the Coulomb potential 
with usual uncertainty $\tilde{E}_n$ to the energy with  generalised uncertainty $E_n$. Here, $M_f=1$TeV, $E_0=13.6$eV is the Rydberg constant. 
\label{fig2}}
\end{figure}

The Coulomb potential is a central potential, hence the virial theorem states that for a particle
moving in this potential, $E_{\rm kin} = - \frac{1}{2}E_{{\rm pot}}$. For an electron of mass $m_{\rm e}$
in the $n$-th level, the total energy $E_n$ is 
\begin{eqnarray}  
E_n = E^{{\rm kin}}_n + E^{{\rm pot}}_n = \frac{1}{2} E^{{\rm pot}}_n = - E^{{\rm kin}}_n\quad.
\end{eqnarray}
Adding the Bohr quantisation condition, the wavelength of the electron fits the circumference of the orbit,
one finds for the $n$th level $\lambda=2\pi n R_n$, hence $k(p) = n/R_n$. Now $E_n = \frac{1}{2} e^2/R_n$, so
the modified $n$-th energy level $\tilde{E}_n$ of the hydrogen atom fulfils
\begin{eqnarray}  
\tilde{E}_n^2 &=& \frac{e^4}{4} \frac{1}{R_n^2} = \frac{e^4}{4} \frac{k(p)^2}{ n^2}
\end{eqnarray}
Inserting now the approximation from eq.~(\ref{app1a}) for $k(p)$, we obtain
\begin{eqnarray}  
\tilde{E}_n^2 &=& \frac{e^4}{4 n^2 \hbar^2} \,p^2 \left(1 - \frac{1}{3}\frac{p^2}{M_f^2}\right)^2\quad.
\end{eqnarray}
Since $\tilde{E}_n = - \tilde{E}^{{\rm kin}}_n = - p^2/2m_{\rm e}$, we can express $p^2$ by $\tilde{E}_n$, which results in the
equation
\begin{eqnarray}  
\tilde{E}_n^2 &=&  - \frac{E_0}{n^2} \tilde{E}_n \left( 1+ \frac{4}{3} \frac{m_{\rm e} \tilde{E}_n}{M_f^2} + \frac{4}{9} \frac{m_{\rm e}^2 \tilde{E}_n^2}{M_f^4} \right) \quad,
\end{eqnarray}
where
\begin{eqnarray}
E_0 = \frac{1}{2} \frac{e^4m_{\rm e}}{\hbar^2} \approx 13.6\,{\rm eV}
\end{eqnarray}
is the Rydberg constant. Introducing the abbreviations
\begin{eqnarray}
\epsilon_n = \frac{E_0}{n^2}\,,\quad \quad \beta = \frac{m_{\rm e}}{3M_f^2}\quad,
\end{eqnarray}
the cubic equation for $\tilde{E}_n$ reads 
\begin{eqnarray}
\tilde{E}_n^2 = -\epsilon_n \tilde{E}_n - 4 \epsilon_n\beta \tilde{E}_n^2 - 4 \epsilon_n \beta^2 \tilde{E}_n^3
\end{eqnarray}
which is solved by
\begin{eqnarray}
\tilde{E}_n 
&=& \frac{1}{8}\frac{1}{\epsilon_n\beta^2}\left(\sqrt{1+8\epsilon_n\beta} - 1 - 4\epsilon_n\beta\right)\quad \nonumber\\
&=& \frac{9}{8}\frac{n^2 M_f^4}{E_0 m_{\rm e}^2}\left(\sqrt{1+\frac{8}{3}\frac{E_0 m_{\rm e}}{n^2 M_f^2}} - 1 - \frac{4}{3}\frac{E_0 m_{\rm e}}{n^2 M_f^2}\right)\quad .
\end{eqnarray}
Neglecting terms of order $O(1/M_f^4) = O(\beta^2)$, an expansion of the square root yields for 
the energy levels the expression
\begin{eqnarray} 
\tilde{E}_n \approx \, - \,\frac{E_0}{n^2} \left( 1 - \frac{4}{3}\frac{m}{M_f^2}\frac{E_0}{n^2}\right) \quad.
\end{eqnarray}
This results fits quite well at higher $n$ to numerical calculations from a perturbative treatment of the 
Hamiltonian we have done. Results are shown in  Fig.~\ref{fig2}.

We can now compare our result with that obtained in \cite{Brau} from perturbation theory.
In that paper it was found that with the modified uncertainty principle, the angular momentum 
degeneracy of the energy levels of the hydrogen atom is lifted. We expect the best match with our
semi classical result for the energy levels of highest angular momentum for a given main quantum number $n$.
In fact, for $l=n-1$, the results of \cite{Brau} exhibit the same dependence of the shift 
on $E_0/n^2$ in the order $O(1/M_f^2)$ for large $n$, differing by a factor $1/3$ from our values. 
We note that the shift found in \cite{Akhoury:2003kc} is similar in size, but has a different sign.
All three results, however, are consistent enough in the absolute value of the shift in the energy levels 
to make comparisons to experimental data. As one might have expected, the deviation caused by the 
modified uncertainty principle is of order $E_0 m_e /M_f^2$, and the $n$ dependence of the shift 
is the same in all three results.

\begin{figure}[h]
\vskip 0mm
\vspace{-0.5cm}
\centerline{
\psfig{figure=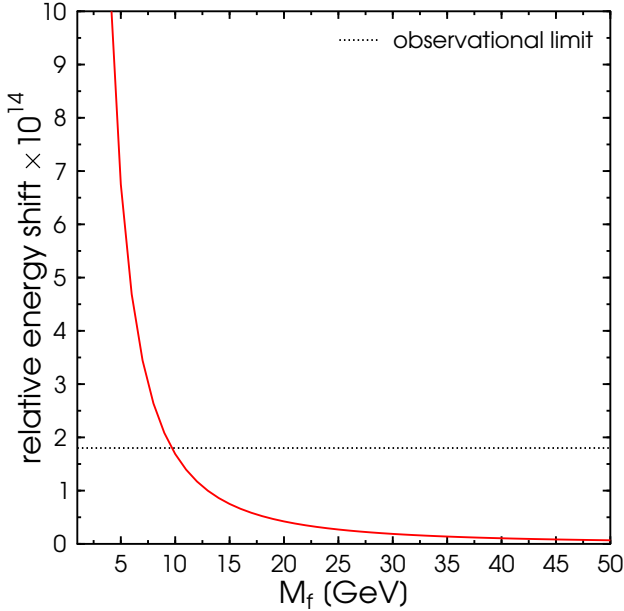,width=4.0in}}
\vskip 2mm
\caption{The relative energy shift of the S1-S2 hydrogen level from usual uncertainty to 
generalised uncertainty as a function of the new scale $M_f$. The horizontal line represents the accuracy of 
todays experiment. Values of energy shift above the observation limit and therefore values 
of $M_f<\approx$ 10 GeV are ruled out.
\label{fig3}}
\end{figure}

To get a connection to experiment, we note that the transition frequency of the hydrogen atom from
S1 to S2 level has been measured up to an accuracy of $1.8\times10^{-14}$ \cite{Niering}. 
In the frequency range of interest, we can certainly neglect transforming the energy into a frequency 
with the new formula. Inserting the values and the current accuracy yields $M_f \approx >50$\,GeV,
as was obtained by \cite{Akhoury:2003kc}. The dependence of relative energy level shift on the
fundamental scale $M_f$ is shown in Fig.~\ref{fig3}, together with the current experimental bound. 
An increase of the experimental precision by four orders of magnitude would 
allow constraints on $M_f$ as tight as the bounds from cosmological and high energy physics. 
An obvious idea would thus be to closely examine constraints arising from high accuracy {\sc QED} predictions, 
such as $g-2$ of the muon and the Lamb shift of the hydrogen atom. 

\section{The Dirac Equation}
In ordinary relativistic quantum mechanics 
the Hamiltonian of the Dirac particle is
\begin{eqnarray}
 \hat{H}= {\rm i} \hbar  \partial_0 = \gamma_0\left( {\rm i} \hbar \gamma^i\partial_i +m \right).
\end{eqnarray} 
This leads to the Dirac Equation
\begin{eqnarray} \label{dirac}
 (p\hspace{-1.5mm}/-m)\psi&=&0,
\end{eqnarray} 
with the following standard abbreviation $\gamma^\nu A_{\nu} := A\hspace{-1.5mm}/$ and $p_{\nu} = {\rm i} \hbar \partial_{\nu}$. 
To include the modifications due to the generalised uncertainty principle, 
we start with 
the relation
\begin{eqnarray} \label{enferm}
 \hat{E} = \gamma_0\left(\gamma^i\hat{p}_i(k) + m \right)
\end{eqnarray} as the first step to quantisation. Including the alternated momentum wave vector relation $\hat{p}(k)$, this yields again Eq. (\ref{dirac}) with the modified momentum operator
\begin{eqnarray} \label{dirac2}
 (p\hspace{-1.5mm}/(k)-m)\psi&=&0 \quad.
\end{eqnarray} 
 This equation is Lorentz invariant by construction (see our general discussion in section \ref{lorentz}). Since it contains -- in position representation -- 3rd order derivatives in space coordinates, it contains 3rd order time-derivatives too. 
In case (a) we can solve the equation for a first order time derivative by using the energy mass shell condition $E^2=p^2+m^2$. This leads effectively to a replacement of time derivatives by space derivatives:
\begin{eqnarray}
\hbar \hat{\omega} \approx \hat{E} - \hat{E}^3/M_f^2 = \hat{E}\left(1- \frac{\hat{p}^i\hat{p}_i + m^2}{M_f^2} \right) \quad .
\end{eqnarray}  
Therefore we obtain the following expression of the Dirac Equation in case (a):
\begin{eqnarray} 
 {\rm i}\hbar    \partial_0 \vert \psi \rangle &\approx&  
\gamma_0\left({\rm i}\hbar \gamma^i\partial_i + m \right) 
\left(1+ L_f^2\partial^i\partial_i - \frac{m^2}{M_f^2} \right) \vert \psi \rangle. 
\end{eqnarray}
However, the equation (\ref{dirac2}) could be considered to be in the more 
aesthetic form, especially --  except in cases we ask for the time evolution -- we will surely prefer its obvious Lorentz invariant appearance.

\section{The Klein Gordon Equation}
Analogously to the derivation of the Dirac Equation in the framework of a
generalised uncertainty principle, we can obtain the modification of the
Klein Gordon Equation.
Again, starting with the energy momentum relation:
\begin{eqnarray}  
  E^2 - p^2 =  m^2, 
\end{eqnarray} 
we obtain
\begin{eqnarray}  
\eta^{\mu\nu} \hat{p}_\nu \hat{p}_\mu  \psi =  m^2 \psi. \label{KGlorentz} 
\end{eqnarray} 
Including the changed momentum wave vector relation $\hat{p}(k)$, 
this yields the
former Klein Gordon equation up to the modified momentum operators.
Note that the square of the generalised Dirac Equation (\ref{dirac2}) 
still fulfils the generalised Klein Gordon Equation (\ref{KGlorentz}).
In case (a), one obtains the following explicit expression in terms of derivative operators:
\begin{eqnarray}  \label{KG}
- \hbar ^2 \eta^{\mu\nu} \left(  \partial_\nu  - \frac{L_f^2}{3} \partial^3_\nu \right) 
\left( \partial_\mu  - \frac{L_f^2}{3}\partial^3_\mu \right) \psi =  m^2 \psi.
\end{eqnarray}

\section{QED}

\subsection{The Fermion Field}
The creation and annihilation operators for anti-particles ${}^- {\hat{a}}^{\dag}_r(p), {}^-\hat{a_r}(p)$ and for particles ${}^+ {\hat{a}}^{\dag}_r(p), {}^+\hat{a}_r(p)$, respectively, obey the following anti commutation relations:
\begin{eqnarray} \label{anticomm}
\left[ {}^+ \hat{a}^{\dag}_r(p) , {}^+ \hat{a}_s(p')\right]_+ &=& 
 \delta_{rs} \delta(\vec{p}-\vec{p}')  \prod_{\nu} \frac{\partial p_{\nu}}{\partial k_{\nu}} \quad ,
 \\
 \left[ {}^- \hat{a}^{\dag}_r(p),{}^-\hat{a}_s(p')\right]_+ &=&  
 \delta_{rs} \delta(\vec{p}-\vec{p}')  \prod_{\nu} \frac{\partial p_{\nu}}{\partial k_{\nu}} \quad.
\end{eqnarray}
and the remaining anti commutators are identically zero. 
The field operator $\psi_{p,r}({\underline {\bf x}})$ can be expanded in terms 
of these creation and annihilation operators in the following way:
\begin{eqnarray} 
\psi_{p,r}({\underline {\bf x}})  &=&  \sum_r \int \frac{{\rm d}^3p}{(2 \pi)^{3/2}}  
\sqrt{\frac{m}{\hbar \omega}}  \; {\rm det} \left( \frac{\partial k_{\nu}}{\partial p_{\mu}}\right) \;  \times\nonumber\\
  \bigg( && \hspace{-5mm} {}^+\hat{a}_r(p) u(p,r) {\rm e}^{{\rm i} k^{\nu}x_{\nu}}  +  {}^-\hat{a}^{\dag}_r(p) v(p,r) {\rm e}^{-{\rm i}k^{\nu}x_{\nu}} 
\bigg).
\end{eqnarray}
In this expression we used the following conventions for the spinors: 
\begin{eqnarray} 
u(p,r) &=& \frac{p\hspace{-1.5mm}/+m}{\sqrt{2m(E+m)}} u(0,r)\quad,  \\
v(p,r) &=& \frac{-p\hspace{-1.5mm}/+m}{\sqrt{2m(E+m)}} v(0,r) \quad,
\end{eqnarray}
where $v(0,r),u(0,r)$ are the unit spinors in the rest frame, $\vec{p}=0$. 
These spinors obey the relations
\begin{eqnarray} 
\sum_r u(p,r)\bar{v}(p,r) &=& \frac{p\hspace{-1.5mm}/+ m }{ 2 m   }  \quad, \\
 \sum_r v(p,r)\bar{v}(p,r) &=& \frac{p\hspace{-1.5mm}/- m}{ 2 m    } \label{end}  \quad.
\end{eqnarray}
 
The Lagrangian density which yields the Dirac equation for the free fermion field is  
\begin{eqnarray} \label{lagfreeferm}
{\cal L}(\bar{\psi},\psi) = {\rm i} \bar{\psi} ( p\hspace{-1.5mm}/(k)-m) \psi \quad. \end{eqnarray}

So we can read off \cite{Ryder,Greiner} the 
free Feynman propagator for the fermions ${}^{\rm f}\Delta_{\mu \nu}$ in momentum representation is
\begin{eqnarray} \label{propfreeferm}
{}^{\rm f}\Delta_{\mu \nu} =   \frac{1}{p\hspace{-1.5mm}/(k) -m + {\rm i} \epsilon} \quad.
\end{eqnarray}
Alternatively, one could have derived this by evaluating time ordered products 
of the field operators using the  relations (\ref{anticomm}) - (\ref{end}).
Evaluating the Feynman propagator by considering these time ordered products yields again (\ref{propfreeferm}) due to the cancellations of the momentum measures by the respective inverse terms.
 
To obtain the Hamiltonian density ${\cal H}$ in the position representation ${\cal H}(x)$, 
one has to treat the Lagrangian density as a function of all appearing 
higher derivative terms ${\cal L}(\psi, \partial_{\nu} \psi, \partial^2_{\mu \nu} \psi, \partial^3_{\mu \nu \kappa} \psi)$ (see \ref{lagfreeferm}). 
Therefore we have to introduce to canonically conjugated momenta:
\begin{eqnarray}
\pi_1(x) &=& \frac{\partial {\cal L}_{{\rm f}}}{\partial(\partial_t \psi(x))} = {\rm i} \psi^{\dag}(x)\quad,\\
\pi_2(x) &=& \frac{\partial {\cal L}_{{\rm f}}}{\partial(\partial^3_t \psi(x))} = - {\rm i} 
\frac{ L_f^2}{3}\psi^{\dag}(x)\quad.
\end{eqnarray}
The Hamiltonian density can now be derived using this generalised scheme:
\begin{eqnarray}
{\cal H}(x)  &=& \pi_1  \partial_t \psi(x)  + \pi_2  \partial^3_t \psi(x)  - {\cal L}\nonumber\\
&=& {\rm i} \psi^{\dag}(x)\partial_t \left( 1 - \frac{ L_f^2}{3} \partial^2_t  \right) \psi(x) \quad.
\end{eqnarray}

\subsection{The Photon Field}
Starting from the expression of the energy density of the photon field in the 
framework of a generalised uncertainty principle:
\begin{eqnarray}
E &=& \frac{1}{4}\tilde{F}^{\mu \nu}  \tilde{F}_{\mu \nu} \quad,
\end{eqnarray}
with the modified field strength tensor, in case (a) explicitly given by
\begin{eqnarray}
\tilde{F}_{\mu \nu} &=& \partial_{\mu} \left( 1 - \frac{ L_f^2}{3} \partial^2_{\mu}  \right) A_{\nu} -
\partial_{\nu} \left( 1 - \frac{ L_f^2}{3}\partial^2_{\nu}  \right) A_{\mu} \quad,
\end{eqnarray}
we derive the corresponding Lagrangian density:
 \begin{eqnarray}
{\cal L} =  - \frac{1}{4}  \tilde{F}^{\mu \nu}  \tilde{F}_{\mu \nu}\quad.
\end{eqnarray}
This can also be expressed as 
\begin{eqnarray}
{\cal L} &=&  - \frac{1}{4} A^{\mu} {\cal{D}}_{\mu \nu} A^{\nu}  
\end{eqnarray}
with
\begin{eqnarray}
{\cal{D}}_{\mu \nu} &=&   \left(  
\stackrel{\leftarrow}{\partial_\mu}-  \frac{ L_f^2}{3}\stackrel{\leftarrow}{\partial_\mu^3}     \right) 
   \left( \stackrel{\rightarrow}{\partial_\nu}  - \frac{ L_f^2}{3}  
  \stackrel{\rightarrow}{\partial_\nu^3}  \right)\quad. 
\end{eqnarray}
Using this Lagrangian density  the interaction-free Feynman photon propagator ${}^{\rm p}\Delta_{\mu \nu}$ 
(in Feynman gauge) can unambiguously determined to be
\begin{eqnarray}
{}^{\rm p}\Delta_{\mu \nu} = - \frac{1}{p^2(k) + {\rm i} \epsilon} \quad.
\end{eqnarray}

\subsection{Coupling}

We introduce the electrodynamical gauge invariant coupling as usual via 
$\partial_{\nu} \to D_{\nu}:= \partial_{\nu} - {\rm i} e A_{\nu}$ in
(\ref{lagfreeferm}). 
We keep as the approximation in case (a) only terms up to first order in $e$ and
terms up to quadratic order in $1/M_f$, admixtures of both are neglected. 
This actually yields the familiar interaction Lagrangian 
\begin{eqnarray}
{\cal L}_{i}  =  -e \bar{\psi} \gamma_{\nu} \psi A^{\nu} \quad.
\end{eqnarray}
As before, we can derive ${\cal H}_{{\rm i}}$, and find as usual
\begin{eqnarray}
{\cal L}_{i}  =  - {\cal H}_{{\rm i}} \quad.
\end{eqnarray}

\subsection{Perturbation Theory}

We see now that the only modification in computing a cross-section 
arises from the different normalisation of the particle states 
and the different volume  factors due to a suppressed occupation of
momentum space at high energies. 
Let us consider now as an important example
the Compton scattering  and ask for the {\sc QED} prediction 
at tree level in perturbation theory.  
We are using the following notation:
\begin{eqnarray}
{\underline {\bf p}}_{{\rm i}1}=(\vec{p}_{{\rm i}1},E_{{\rm i}1}) &:&
\mbox{initial electron} \quad,
\nonumber\\
{\underline {\bf p}}_{{\rm i}2}=(\vec{p}_{{\rm i}2},E_{{\rm i}2}) &:&
\mbox{initial photon}  \quad,
\nonumber\\
{\underline {\bf p}}_{{\rm f}1}=(\vec{p}_{{\rm f}1},E_{{\rm f}1}) &:&
\mbox{final electron} \quad,
\nonumber\\
{\underline {\bf p}}_{{\rm f}2}=(\vec{p}_{{\rm f}2},E_{{\rm f}2}) &:&
\mbox{final photon} \quad,
\end{eqnarray}
and
 \begin{eqnarray}
 {\underline {\bf p}}_{{\rm i}}&=&{\underline {\bf p}}_{{\rm i}1}+{\underline {\bf p}}_{{\rm i}2} \quad, \nonumber\\
 {\underline {\bf p}}_{{\rm f}}&=&{\underline {\bf p}}_{{\rm f}1}+{\underline {\bf p}}_{{\rm f}2} \quad, \nonumber\\
 E&=&E_{{\rm i}1}+E_{{\rm i}2}=E_{{\rm f}1}+E_{{\rm f}2}
 \quad.
 \end{eqnarray}
The expression of the $S$-Matrix element in the realm of the generalised uncertainty principle is:
\begin{eqnarray}
\tilde{S}_{{\rm fi}} = (2 \pi)^4 \tilde{M}_{{\rm fi}} \delta( {\underline {\bf p}}_{{\rm i}} -  {\underline {\bf p}}_{{\rm f}})  \prod_{\nu} \frac{\partial p_{\nu}}{\partial k_{\nu}}\Bigg|_{ {\underline {\bf p}}_{{\rm i}}={\underline {\bf p}}_{{\rm f}}}
\end{eqnarray}
The probability of the initial particles to wind up in a certain range of momentum space d$P($i$\to$ f$)$ can
be obtained in the usual way by putting the system into a finite box with volume $V$. 
Since the measure of momentum space is modified this yields a Jacobian determinant for every final particle. For our example this reads
\begin{eqnarray}
{\rm d}P ({\rm i}\to {\rm f}) = \left( \frac{(2 \pi)^3}{V} \right)^2 \vert \tilde{S}_{{\rm fi}} \vert^2 
\hbar^2 {\rm d}^3p_{\rm f1} {\rm d}^3p_{\rm f2} \left( \prod_{\nu} \frac{\partial k_{\nu}}{\partial p_{\nu}}\right)^2
\end{eqnarray}
and the differential cross-section for two particles in the final state is then
\begin{eqnarray}
{\mathrm d} \tilde{\sigma} \left({\rm i} \to {\rm f}\right) = \hbar^2 (2\pi)^4 \frac{1}{\Phi V} \vert \tilde{M}_{{\rm fi}} \vert^2 
\frac{ E_{{\rm f}1} E_{{\rm f}2} \vert p_{{\rm f}2} \vert}{E}  \prod_{\nu} \frac{\partial k_{\nu}}{\partial p_{\nu}} {\mathrm d} \Omega \quad.
\end{eqnarray} 
Here $\Phi$ is the flux. In the laboratory system, we have $\vec{p}_{{\rm i}1}=0, E_{{\rm i}1}=m, E_{{\rm f}1}=m+E_{{\rm i}2}-E_{{\rm f}2}$, therefore
$\Phi V=1 $. This leads to the following expression in the laboratory system:
\begin{eqnarray} \label{crosstwopart}
{\mathrm d} \tilde{\sigma} \left({\rm i} \to {\rm f}\right) = \hbar^2(2\pi)^4   \vert \tilde{M}_{{\rm fi}} \vert^2 
\frac{ E_{{\rm f}1} E_{{\rm f}2}^3  }{m E_{{\rm i}2}}  \prod_{\nu} \frac{\partial k_{\nu}}{\partial p_{\nu}} {\mathrm d} \Omega \quad.
\end{eqnarray}
Explicitly, in case (a) and in the laboratory system, we have
\begin{eqnarray}
\prod_{\nu} \frac{\partial p_{\nu}}{\partial k_{\nu}} = \hbar^4
\left(1+ \frac{E_{{\rm i}2}^2}{M_f^2}\right)\left(1+ \frac{(m+E_{{\rm i}2})^2}{M_f^2}\right)
\end{eqnarray}
and the Jacobi determinant of the inverse function in Eq. (\ref{crosstwopart}) is just given by the inverse of this expression.

The amplitude summed over all possible initial and final polarisations, $e_{{\rm i}}$, $e_{{\rm f}}$,
remains in the well known standard form
\cite{Weinberg}
\begin{eqnarray}
\frac{1}{4}\sum_{ {\sigma}_{{\rm i}} \sigma_{{\rm f}} e_{{\rm i}} e_{{\rm f}} } 
\vert \tilde{M}_{{\rm fi}} \vert &=& \frac{e^4}{64 (2 \pi)^6} 
\frac{1}{\omega_{{\rm f}2} \omega_{{\rm i}2} \omega_{{\rm f}1} \omega_{{\rm i}1}} \;\times\nonumber\\
&&\left[ \frac{E_{{\rm i}2}}{E_{{\rm f}2}} + \frac{E_{{\rm f}2}}{E_{{\rm i}2}} -1 + \cos^2\theta\right]
\end{eqnarray}
with $\omega_{\rm index}=\omega(E_{\rm index})$. 
All this put together yields
\begin{eqnarray}
\frac{1}{4}\sum_{\sigma_{{\rm i}} \sigma_{{\rm f}} e_{{\rm i}} e_{{\rm f}} }  
\frac{{\mathrm d} \tilde {\sigma}}{{\mathrm d} \Omega} \left({\rm i} \to {\rm f}\right) &=&
\frac{\hbar^2 e^4}{32   \pi^2} \prod_{\nu} \frac{\partial k_{\nu}}{\partial p_{\nu}} \; \times \nonumber\\
\frac{(m+E_{{\rm i}2}-E_{{\rm f}2})E_{{\rm f}2}^3}{\omega_{{\rm f}2} \omega_{{\rm i}2} \omega_{{\rm f}1} \omega_{{\rm i}1} m E_{{\rm i}2}}  
&&\left[ \frac{E_{{\rm i}2}}{E_{{\rm f}2}} + \frac{E_{{\rm f}2}}{E_{{\rm i}2}} -1 + \cos^2\theta\right]
\quad.
\end{eqnarray}
This example illustrates how modified cross sections $\tilde{\sigma}$ 
in scattering processes with two initial and two final states can be obtained 
from the unmodified cross sections $\sigma$.
This relation is given by the following formula:
\begin{eqnarray} \label{dsds}
\frac{{\mathrm d} \tilde {\sigma}}{{\mathrm d} \sigma} = \prod_{n} \frac{E_n}{\omega_n}  \prod_{\nu} \frac{\partial k_{\nu}}{\partial p_{\nu}}\Bigg|_{ {\underline {\bf p}}_{{\rm i}}={\underline {\bf p}}_{{\rm f}}}
\quad.
\end{eqnarray}
From the steps of calculation it can be seen that 
this result holds in higher order perturbation
theory, too. 
The modification 
enters through the energies of the in- and outgoing particles and their
momenta spaces, only. 
However, when incorporating higher orders one has to bear in mind, 
that we approximated the interaction Hamiltonian 
by neglecting terms of order $\alpha/M_f^2$. 
These terms should reappear at higher energies leading to the necessity of
a reordering of the corresponding perturbation series. To be precise, 
the full modified SM result contains more terms than one would have taken into
account by just using Eq. (\ref{dsds}).

Let us interpret this result physically before going any further. 
There are two factors occurring. The first shows that the physics 
at a certain energy $\tilde{E}$ of two particles 
is now rescaled. It is identical to the physics that 
happened before at a smaller energy $E$ with $E=\hbar\omega(\tilde{E})$.
A higher energy is needed within our model 
to reach the same distance between the particles as in the standard model:
To get the same resolution as with the standard uncertainty principle, 
one has to go to higher energies! 
Because the cross sections decrease with energy this means 
our modified cross section predictions are higher at the same energy than those of
the standard model. 
The functional behaviour of the
standard model result should be cut at $M_f$ 
and the range up to $M_f$ be stretched out to infinity.
In particular, only from this factor the cross-section would asymptotically 
get constant at a value equal to the unmodified standard model result at $M_f$. 

But there is another factor from the Jacobian, which takes into account 
that the phase space for the final states is reduced significantly from Planckian energies on.
Since ${\bf{\underline k}}({\bf{\underline p}})$ approaches a constant value, 
its Jacobian and therefore the relation ($\ref{dsds}$) drops to zero. 
Putting  both effects together, the cross section of our model drops below 
the unmodified standard model result:  
As can be seen from the Jacobian in case (b), the 
cross section drops exponentially with the reaction energy. 
 
The prediction of a dropping cross section in comparison to the unmodified 
standard model results is quite remarkable. In most models with the assumption 
of extra dimensions only, an increase
of the cross section is predicted.\footnote{Note: \cite{Giudice:1998ck} mentions the 
possibility of a dropping cross section in the realm of large extra dimension 
scenarios.} 
This is due to the enhanced possible reactions when taking
into account virtual gravitons (see next section). 

It is obvious by 
construction that in our model
no physics can be tested below the distance $L_f$. 
If the new scale is as low as $\approx$ TeV, 
as suggested by the proposal of Large Extra Dimensions, 
then an even further increase of the 
energy that can be delivered by even larger 
colliders than the next generation can deliver ($\approx$ 14 TeV at {\sc LHC}) would 
not yield more insight than the statement that there is such a
smallest scale in nature. As was formulated by Giddings  
this would be ``the end of short distance physics'' \cite{Banks:1999gd,Giddings:2001bu}.
However, this was mentioned in a different context. 
In our approach the production of tiny black holes is not yet possible 
at center of mass (c.o.m.) energies $\sqrt{s} \approx M_f$, because 
the distance needed for two partons of energy $\approx M_f$ 
to collapse and form a black hole is just $L_f$, but the particles 
can not get that close any longer. (This might happen then at higher energies, see 
\cite{Gross}.) Therefore, we are most interested in testing the present model
in ultra high energetic cosmic ray experiments, like the extended
air-shower measurements at KASCADE-Grande and at 
the Pierre Auger Observatory \cite{BLUB!}, which allows
 a hundredfold c.o.m.-energy increase over the {\sc LHC} energies.

For energy $\sqrt{s}$ ($m_{\tau},m_{\mu}\ll\sqrt{s}$) ,
Eq. (\ref{dsds}) yields the simple expression with the functions inserted in the c.o.m. system
\hfill

\begin{eqnarray}
\frac{{\mathrm d} \tilde {\sigma}}{{\mathrm d} \sigma} &=&  \frac{s^2}{(2 M_f)^4} 
\left( \tanh\left(\frac{\sqrt{s}}{2 M_f}\right)\right)^{-4} \times
\nonumber \\ 
&&\left( \cosh^2\left(\frac{\sqrt{s}}{ M_f}\right)\right)^{-2}.
\end{eqnarray}

\begin{figure}[h]
\vskip 0mm
\vspace{0cm}
\centerline{
\psfig{figure=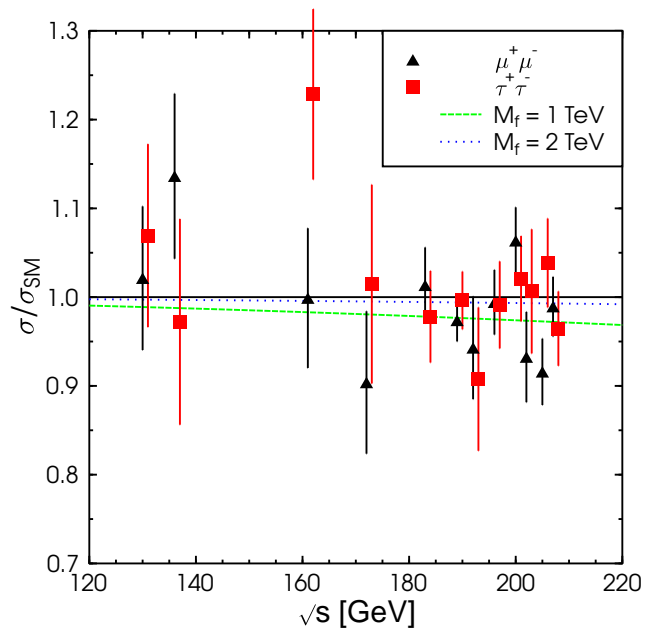,width=4.0in}}
\vskip 2mm
\caption{The ratio of $e^+e^-\to\mu^+\mu^-, \tau^+\tau^-$ cross sections
calculated with the generalised and the ordinary uncertainty principle.
Data is taken from \protect\cite{lep}.
\label{fig4}}
\end{figure}

We have used this functional behaviour to get the connection to the measured 
data of the LEP2 collaboration, \cite{lep}, $e^+e^-\to\mu^+\mu^-$ and  $e^+e^-\to\tau^+\tau^-$ cross-sections. The derived factor 
is independent of the scattering angle. Hence, it holds for the total cross-section, 
as shown in Fig \ref{fig4}. In this context, 
note that possible limits on physics beyond the standard
model in LEP2 fermion pair production data
have already been discussed in \cite{Abbiendi} from 
the experimental view -- this is one of the new trends in 
high-energy physics.
 
\section{Gravitons}
 
Many prominently discussed collider signatures of {\sc LXD}s are connected to the 
virtual and real graviton $G$ production processes. 
Extensive studies of this subject already
exist in the literature (see e.g. \cite{Giudice:1998ck,Hewett:2002hv,Han:1998sg}).
In these scenarios Kaluza-Klein excitations are given in steps $n_a/R$.
The maximum possible frequency $n_a=R M_f$ is in these scenarios 
the natural cutoff. (For simplicity we have set the compactification radii of all extra dimensions to be equal.)

We start with the real gravitons, which are important at energies $>M_f$
due to the significant increase of the corresponding phase space factor.
In order to estimate the $e^+e^-\to \gamma G$ cross section in the context
of the modified uncertainty principle, we start with the relation:
\begin{eqnarray}
\sigma(e^+e^-\to \gamma G) \propto   \frac{e^2}{m_p^2} N(\sqrt{s})\quad,
\end{eqnarray}
we have to calculate the number of possible final states with $E=\sqrt{s}/2$ in the c.o.m. system,
which is called $N(\sqrt{s}/2)$.
$N(\sqrt{s}/2)$ can be obtained using (with $\Delta m \rightarrow {\rm d}m$):
 \begin{eqnarray}
\frac{{\rm d}m}{{\rm d}n_a} = \frac{1}{R} \frac{\partial E}{\partial \omega} \quad,
\end{eqnarray}
where $m$ is the apparent mass of the excitation of the respective Kaluza-.Klein state:
\begin{eqnarray}
m^2= p_\perp^2 = \sum_{a=4}^{d+4} \left(E(\omega_a)\right)^2\;, \hspace{0.3cm}
{\rm with}\quad
\omega_a =
\frac{n_a}{R}\quad.
\end{eqnarray}

Using  Eq. (\ref{master}) and the above expressions one obtains for the number of final states:
\begin{eqnarray}
N (\sqrt{s}) &=& \Omega_{(d-1)} R^{d} \int_0^{\sqrt{s}/2} {\rm d} m \; \omega(m )^{d-1} \frac{\partial \omega}{\partial E} \nonumber \\
&=& V_{(d)} \frac{m_p^2}{M_f^{d+2}}  \omega(\sqrt{s}/2 )^{d}
\end{eqnarray}
with $\Omega_{(d-1)}$ being the surface of the $d$-dimensional unit-sphere and 
$V_{(d)}$ being its volume:
\begin{eqnarray}
\Omega_{(d-1)} &=& \frac{2 \pi^{d/2}}{\Gamma\left(d/2\right)} = d V_{(d)} \quad.
\end{eqnarray}
 These considerations yield the following estimation of the real graviton production cross section:
\begin{eqnarray} \label{crossest}
\sigma(e^+e^-\to \gamma G) \propto \frac{e^2}{\omega(\sqrt{s}/2)^2} \left( \frac{\omega(\sqrt{s}/2)}{M_f} \right)^{d+2}.   
\end{eqnarray} 
The exact result for the fermion to real graviton plus $\gamma$ cross section  
in the framework of the generalised uncertainty principle depends on the amplitude 
of the process and on the spin-sums. However, for the following general considerations the estimate (\ref{crossest}) is sufficient.  
This cross-section would be of the same importance 
with SM processes if $\omega$ equals $M_f$, which is
here only possible asymptotically. Therefore, real gravitons 
are produced at a lower rate when a generalised uncertainty principle is employed 
than expected from LXD scenarios without the generalised uncertainty relation. 
As a consequence,  
constraints (e.g. by energy loss) from real graviton emission should be reanalysed carefully
in the context of the minimal length proposal.

Now, let us turn to the virtual graviton production. The 
free graviton propagator from  \cite{Giudice:1998ck} for $G_m$ (Graviton of apparent mass $m$) is generalised to:
\begin{eqnarray}
{}^{\rm G}\Delta=\frac{P_{\mu \nu \alpha \beta}}{p(k)^2-m^2} \quad,
\end{eqnarray} 
where $P_{\mu \nu \alpha \beta}$ is the graviton polarisation tensor 
(the exact form of the polarisation tensor can be found in \cite{Giudice:1998ck}).
To calculate the complete graviton exchange amplitudes, the amplitudes for different $m$ have to be summed up.
The ultraviolet-divergence of this sum has to be fixed 
by introducing a cut-off parameter $\Lambda$ that is of order $M_f$.
Such an ad hoc introduction of a cut-off parameter 
is from a theoretical point of view always somewhat dissatisfying.
In the context of the generalised uncertainty relation such a cut-off parameter
is naturally included from first principals via the minimal length scale $L_f$.
Therefore, no ad hoc cut-off parameter is needed:
\begin{eqnarray}
\sum_n\frac{1}{p^2-m^2} = \Omega_{(d-1)}\int_0^{\infty} {\rm d} m \frac{m^{d+1}}{p^2-m^2} \frac{\partial \omega}{\partial E}.
\end{eqnarray} 
Using case (b), it is easy to see that the UV-end converges for all $d$ due to the 
exponential suppression of the momentum measure. 
To calculate this integral, it can be expanded in a power series 
in $\sqrt{s}/M_f$,  as given in \cite{Giudice:1998ck} using the cut-off parameter. 
In our approach the expansion coefficients could be calculated right away. 
We will not perform this analysis here. This result 
will not yield a more profound relation between the exact parameters 
and the expansion coefficients, since 
in our approach the arbitrariness lies in the exact 
form of the   function $E(\omega)$ applied,
or its expansion coefficients, respectively.

Even if the details of graviton production are not 
further examined in this paper, one can now
conclude that within our model the 
cross-sections 
(e.g. the above calculated $\tilde{\sigma}(e^+e^-\to f^+f^-)$) are 
modified in a different way than in the scenario with 
{\sc LXD}s only. The virtual graviton
exchange increases the cross-section, but the squeezing of the 
momentum space decreases it. So we
have two effects of a similar magnitude 
which are working against each other. 
Therefore, measurable deviations may occur only at  energies higher 
than $M_f$. If one is looking for signatures
beyond the standard model, one should focus instead on observables 
that are not too sensitive to the
generalised uncertainty, such as modifications in the spin distribution 
due to the exchange of a
spin 2- particle  or the
 appearance of processes that are forbidden by the standard-model.
Furthermore, we want to mention that most of the constraints on the 
$M_f$ scale are weakened in our scenario.

\section{Conclusions}

We introduce modifications of quantum mechanics caused by the
existence of a minimal length scale $L_f$. We show that our approach is
consistent with other calculations on this topic. Assuming the recent
proposition of Large Extra Dimensions, the new scale might be accessible 
in colliders.
We use perturbation theory to derive the $e^+e^- \to f^+f^-$ cross sections
with an approximated interaction Hamiltonian. We compare
our results to recent data and find that the limits on the new
scale are compatible to those from different experimental constraints: 
$1/L_f \approx>1$ TeV. Our model combines both
Large Extra Dimensions and the minimal length scale and predicts dropping 
cross sections relative to the standard model cross sections. Further, we argue that 
the analysed Planckian effects hinder the emergence of other effects which are
predicted above $\approx 1$ TeV, such as black hole and graviton production.

\section*{Acknowledgements}
The authors thank L.~Bergstr\"om, S.~F.~Hassan 
and A.~A.~Zheltukhin for fruitful discussions.
S.~Hofmann and S.~Hossenfelder appreciate the kind hospitality
of the Field and Particle Theory group at Stockholm University.
S.~Hofmann acknowledges financial support from the
Wenner-Gren Foundation.
S.~Hossenfelder wants to thank the Land Hessen for financial support.
J.~Ruppert acknowledges support from the 
Studienstiftung des deutschen Volkes (German national merit foundation).
This work has been supported by the {\sc BMBF}, {\sc GSI} and {\sc DFG}.

\end{document}